\documentclass[a4paper,11pt]{article}

\usepackage{listings}
\usepackage{subcaption}
\usepackage{xcolor}
\usepackage{graphicx} 
\usepackage{url}
\usepackage{booktabs} 

\usepackage[left=2.5cm, right=2.5cm, top=2.5cm, bottom=2.5cm]{geometry}

\definecolor{codegreen}{rgb}{0,0.6,0}
\definecolor{codegray}{rgb}{0.5,0.5,0.5}
\definecolor{codepurple}{rgb}{0.58,0,0.82}
\definecolor{backcolour}{rgb}{0.95,0.95,0.92}

\lstdefinestyle{mystyle}{
    commentstyle=\color{codegreen},
    keywordstyle=\color{magenta},
    numberstyle=\tiny\color{codegray},
    stringstyle=\color{codepurple},
    basicstyle=\ttfamily\footnotesize,
    breakatwhitespace=false,         
    breaklines=true,                 
    captionpos=b,                    
    keepspaces=true,                 
    numbers=left,                    
    numbersep=5pt,                  
    showspaces=false,                
    showstringspaces=false,
    showtabs=false,                  
    tabsize=2
}

\title{The enshittification of online search?\\Privacy and quality of Google, Bing and Apple in coding advice\\[1ex] \large Technical report: non-peer-reviewed of work in progress}
\author{Konrad Kollnig\\ \textit{Law \& Tech Lab, Maastricht University, The Netherlands}}
\date{Last update: December 2025}

\begin{document}

\maketitle

\begin{abstract}
\noindent
Even though currently being challenged by ChatGPT and other large-language models (LLMs), Google Search remains one of the primary means for many individuals to find information on the internet.
Interestingly, the way that we retrieve information on the web has hardly changed ever since Google was established in 1998, raising concerns as to Google's dominance in search and lack of competition.
If the market for search was sufficiently competitive, then we should probably see a steady increase in search quality over time as well as alternative approaches to the Google's approach to search.
However, hardly any research has so far looked at search quality, which is a key facet of a competitive market, especially not over time.

In this report, we conducted a relatively large-scale quantitative comparison of search quality of 1,467 search queries relating to coding advice in October 2023. We focus on coding advice because the study of general search quality is difficult, with the aim of learning more about the assessment of search quality and motivating follow-up research into this important topic. We evaluate the search quality of Google Search, Microsoft Bing, and Apple Search, with a special emphasis on Apple Search, a widely used search engine that has never been explored in previous research.
For the assessment of search quality, we use two independent metrics of search quality: 1) the number of trackers on the first search result, as a measure of privacy in web search, and 2) the average rank of the first Stack Overflow search result, under the assumption that Stack Overflow gives the best coding advice.
Our results suggest that the privacy of search results is higher on Bing than on Google and Apple.
Similarly, the quality of coding advice – as measured by the average rank of Stack Overflow – was highest on Bing.

Recognising the lack of reliable and longitudinal assessments of search quality, we propose to develop new methodologies in this area and discuss the implications of new laws, notably the EU Digital Markets Act (DMA).
Crucially, as more individuals shift to LLMs which are arguably even more opaque than traditional search, an assessment of search/information quality will be ever more important.
\end{abstract}


\section{Introduction}
In the years following the launch of the first website by Sir Tim Berners-Lee at CERN on 6 August 1991~\cite{bernerslee_worldwide_1992}, the size and dimension of the World Wide Web (WWW) was first created was nimble at first.
Only a few of us even had access, mostly those based in university laboratories, and much of the information flows were still happen via more traditional means such as telephone or (electronic) mail.
With personal computing revolution, this, however, quickly started to change once more individuals owned computers and were connected to the internet.
While only about 14\% of US households had internet access in 1995, this figure had more than tripled~--~to 46\%~--~by the year 2000~\cite{internetuse}.

With this exponential adoption of computing technology and the internet also came an exponential increase in the amount of available information.
At first, this information was organised through manually curated directories on websites. Yahoo!, founded in 1994 by Jerry Yang and David Filo, started as one such directory. Yahoo!'s approach was to manually categorise websites into a hierarchical directory.
However, manually curating and categorising the web quickly became impractical due to its rapid expansion. This led to the development of automated search engines. Early examples include AltaVista (1995), which was notable for its fast, multi-threaded crawler, and Lycos, which started as a research project at Carnegie Mellon University.

Today's behemoth in online search, Google, was only founded in 1998 by Larry Page and Sergey Brin. It mangaged to revolutionise search engine technology by using a new algorithm based on website relevance and backlinks (PageRank)~\cite{brin_anatomy_1998}. This was more effective than the keyword-based approach of earlier search engines.
Over time, some early search engines failed to adapt and fell behind. For instance, AltaVista lost ground due to its cluttered interface and the shift of focus by its parent companies. Lycos and Excite also lost relevance with the rise of more efficient search engines like Google.
Nowadays, Google is the most widely used search engine, with a market share exceeding 90\% of internet users worldwide; Bing, its closest competitor, only has a share of 3\%~\cite{search_share}.

Google's high market share underlines the success of its model that manages to generate extremely high revenues from a tight integration of its core service (Search) with its other services (e.g. YouTube, Chrome, Drive, Docs).
The vast majority of Google's revenue~--~which reached 282 billion USD in 2022~--~is derived from online advertising, which makes up more than 80\% of its overall revenues~\cite{google_ad_revenue}.
Importantly, Google does not only retrieve advertising revenue from showing advertising on its own services, but also on third-party apps and websites.
This is because Google is the most dominant company across all online advertising outside of social media~\cite{cma_digital_markets2020}.

Google's dominance across these services also creates a variety of potential concerns. By operating across many segments of digital technologies, Google often has a competitive advantage over competitors from its cross-market dominance~\cite{khan_amazons_2017,morton_roadmap_nodate}.
For example, it is not clear if Google's dominance in search is derived from its innovative nature or rather from the fact that it has been the first company to take up that market (`market capture').
In the economics literature, two important indicators of a competitive market are improvement in product price or quality over time~\cite{wu_curse_2018}.
Given that Google Search comes without a price, the study of search quality might be the more tractable problem for the research community.
If the market around search is sufficiently competitive, then one might expect that search quality has steadily improved over time.
Despite that, hardly any previous research has attempted to assess and monitor search quality.

\begin{figure}
    \centering
    \includegraphics[width=0.4\columnwidth]{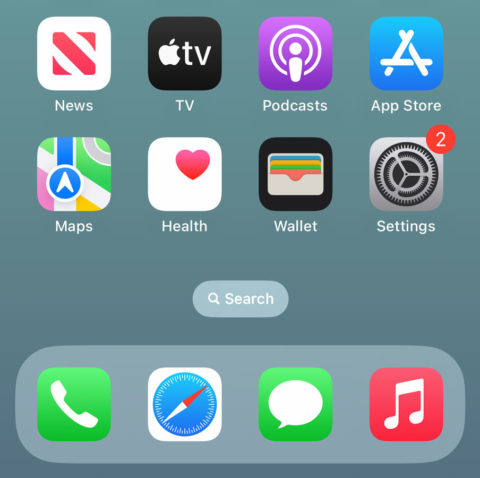}
    \caption{iOS 16, released in September 2022, brought Apple Search to the home screen and made it more prominent than ever before. Apple Search only covers searches from within Spotlight search but not the Safari browser. At the moment, Apple has limited incentive to deploy its search engine widely in Safari because it receives major payments from Google for selecting its Search as the default in the Safari browser. The company is, however, arguably getting prepared to move away from using Google Search as a default, especially if Google should be forced to stop payments to Apple following the ongoing US court proceedings. Apple Search often gets overlooked in the debate around online search, even though it is used by hundreds of millions of Apple users every day.}
    \label{fig:apple-s}
\end{figure}

In the interim, we have arguably been left with an approach to information retrieval, namely online search, that has hardly changed since Google's launch in 1998.
This is, admittedly, slowly changing with the wide adoption of LLMs~\cite{stackoverflow-traffic} but it remains unclear whether to what extent they are going to innovate within search.
Hence, this report focuses on the study of search quality in traditional search over time.
To this end, we study the case study of coding advice in October 2023, being one of the largest academic studies done on the subject of search quality (comparing more than 1,000 search queries across popular search engines). From this, we abstract how search quality might be continuously assessed over time.
In terms of search engines, we focus on the most popular search engines, namely Google Search, Microsoft Bing, and Apple Search. The latter is part of every iOS / macOS device but rarely noticed by users and also rarely studied, see Figure~\ref{fig:apple-s}.

\section{Past work on Search Quality}

The past work on search quality is somewhat limited and falls into two main categories: industry self-studies, and old or small-scale academic studies. Both usually rely on some form of human evaluation of search results.

The industry self-studies, mainly by Google and Bing, used qualitative approaches such as interviews and surveys so as to understand users' levels of satisfaction with search as well as their preferences for specific search engines.
For example, Google regularly conducts `User Satisfaction Tests', in which individuals rate unbranded search results.
In these tests, Google tends to outperform Bing in all categories~\cite{cma-search-quality}.
While qualitative research methods can be important, they are expensive to conduct repeatedly, difficult to scale, and might suffer from lack of objectivity.
There are also concerns around the neutrality of Google and Bing conducting such studies.
This makes it difficult to interpret these self-studies and use them as definite evidence.

In terms of academic studies, much previous work on search quality looked at a small set of search queries, typically under 100, comparing the results between various search engines, and using human evaluation as a quality measure~\cite{goutam_performance_2012,shafi_retrieval_2019,hawking_measuring_2001,vaughan_new_2004}.
Some of the largest work stems from Lewandowski and evaluated 1,000 informational and 1,000 navigational queries in German~\cite{lewandowski_evaluating_2015}.
Google was found to outperform Bing on informational queries, albeit marginally.
For navigational queries, this margin was higher, with Google leading the correct result in 95.3\% of searches compared to 76.6\% for Bing.
However, this study was done with data from 2011 and relied on human judgement, which is costly and can lack objectivity similar to the industry reports.

Overall, we currently lack robust methods to study search quality over time and at an affordable cost for researchers.
Furthermore, Apple Search has never been studied before.
Hence, this work tries to make important contributions towards building more scalable and objective indicators of search quality than previous work.
It should be noted that we only focus on a case study and that more work needs to be done.

\section{The Evolution of Google Search and Role of Search Engine Optimisation (SEO)}

Google Search has seen various stages of development of time. While the idea behind PageRank remains relevant even today, the company had to make various adjustments over time to cope with changing trends and with individuals trying to game the system.
Interestingly, most of these major updates only happened after the last major study into search quality~\cite{lewandowski_evaluating_2015}, which used data from 2011, further underlining the need for renewed and continuous, analysis.

Panda, from 2011, was Google's first major publicly announced update to its search. This update targeted sites with low-quality content. Google's aim was to lower the rank of websites with poor-quality content and elevate higher-quality sites. This move encouraged webmasters to focus on creating meaningful, original content rather than relying on large quantities of low-quality material. Google's 2012 Penguin update further sought to penalise websites engaging in manipulative link practices. It devalued spam and irrelevant links, making practices like link farming and the purchase of backlinks less effective, thereby addressing the potential abuse of PageRank.

`Mobilegeddon' (as termed by the industry at the time) was another significant step in 2015. With the increasing prevalence of mobile internet usage, this update now prioritised mobile-friendly websites in Google's mobile search results. Websites that were not optimised for mobile devices saw a decrease in mobile search rankings, pushing webmasters to adopt responsive or mobile-specific site designs.
The shift towards being a mobile-first search engine took many years is still being worked on.

With the steady improvement in machine learning, Google sought to integrate those developments into its search. RankBrain, from 2015, introduced various machine learning techniques into Google's search algorithm. This component helped Google to process search results and provided a better semantic understanding of user intent, especially for complex or novel queries.
The arguably most recent significant innovation was, however, Google's invention and release of Bidirectional Encoder Representations from Transformers (BERT) in 2018~\cite{devlin_bert_2019} and its subsequent integration into Google Search.
BERT is the theoretical foundation for all current LLMs and helped Google better understand the context of each word in a search query~--~besides now being severely challenged in its core business.

Each of these steps in the evolution of Google Search were in response to the development of new technologies, to companies trying to cheat Google's system, and to companies' needs to understand how they can reach their customers with their content.
The targeting of websites' content to Google's system is often called Search Engine Optimisation (SEO) and has a somewhat mixed standing.
On the one hand, it is positive that Google publicly documents the parameters that it uses to rank its search results.
On the other hand, SEO, in the past, has encouraged a focus on search rank rather than content quality, which underlines the challenges that Google faces in ranking content fairly and transparently.

\section{Methods}
\label{sec:methods}

We study the search quality on the most popular search engines (Google Search, Microsoft Bing, Apple Search), focusing on the case study of coding advice.
In this study, \textbf{we \textit{assume} that Stack Overflow tends to give the best coding advice when it comes to common Python coding questions}, and should thus rank high in the search results that pertain to common Python coding questions.
Stack Overflow is a popular online forum, in which users can post coding-related questions, give answers to other users' questions, and vote for the best answers.
In this way, over the years, there has emerged a relatively full coverage of common Python coding questions.
As of December 2023, Stack Overflow had more than 22 million registered users, 24 million questions, and 35 million answers~\cite{stackoverflow-stats}.

\begin{figure}
    \centering
    \includegraphics[width=0.7\columnwidth]{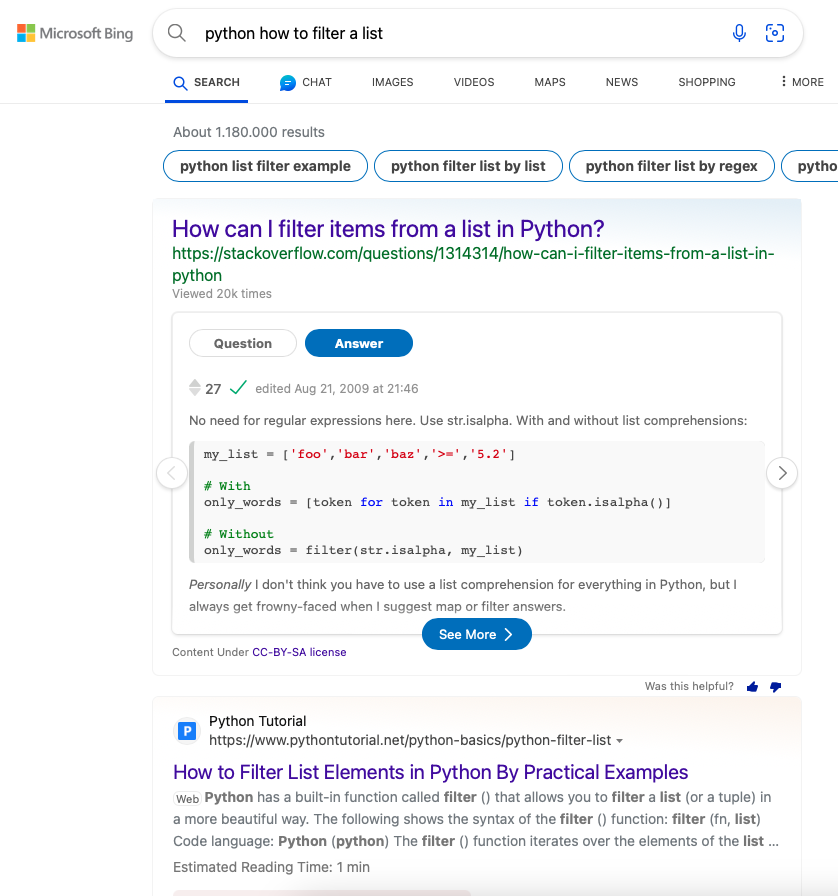}
    \caption{Compared to other search engines, Bing often embeds the code snippets from Stack Overflow directly into the search results. This suggests that developers of Bing made some intentional decisions to implement such an embedding and that they agree with our assessment that Stack Overflow regularly gives the best coding advice.}
    \label{fig:bing-stackoveflow}
\end{figure}

We admit that this is an imperfect assumption but believe it is a fair approximation, given that past academic research has tended to look at 10--20 search queries or is very dated. With our approach, we managed to compare more than 1,000 queries across popular browsers.
Also, Microsoft seems to agree with our analysis, given that they regularly embed code snippets from Stack Overflow directly into the Bing search results, see Figure~\ref{fig:bing-stackoveflow}.
To do so, the company most likely made an intentional effort to give prominence to Stack Overflow when it comes to coding-related questions, so as to increase search quality.
Furthermore, Stack Overflow is in many ways what Wikipedia is for general knowledge; due to a a large number of volunteers, such platforms easily become the number one resource for information in their respective fields~\cite{benkler_wealth_2006}.

\begin{figure}
    \centering
    \begin{verbatim}
        python how to download a file from url
        python how to flatten a list
        python how to check string is empty
        python how to read txt file
        python how to round down
        python how to view installed packages
        python how to mock a function
        python how to get size of file
        python how to end a while loop
        python how to populate a list
        python how to remove item from list
    \end{verbatim}
    \caption{An excerpt from the list of generated search queries for coding advice.}
    \label{fig:suggestions}
\end{figure}

In order to determine a list of common Python questions, we fed the autocomplete functionality of Google Search, which available under the endpoint \url{http://suggestqueries.google.com/complete/search?output=toolbar&hl=en&q=[query]}, with queries starting with `python how to ', followed by all two letter words consisting of the 26 letters of the roman alphabet.
Hence, we search for `python how to aa', `python how to ab', `python how to ac', etc.
Excluding duplicates, this resulted in a total of 1,467 unique coding questions, see Figure~\ref{fig:suggestions}.
Given that the search suggestions on Google Search are based on the most common search queries, this gives an approximate list of common coding questions in Python.
The first author sense-checked the search queries so as to make sure that they actually contain coding-related questions.

Following up, we fed all the 1,467 Python coding questions into Google Search, Bing and Apple Search.
For Google Search and Bing, we used puppeteer~\cite{puppeteer}, which provides an automation interface to Google Chrome, and extracted all the search results on the first results page.
In this way, we were able to overcome rates limits with the unofficial API endpoints related to that search.
Given that Apple Search does not have a web front-end, we reverse-engineered its API by using Charles Proxy~\cite{charles} on a MacBook running the latest version of macOS.
This is visualised and documented in Figure~\ref{fig:apple-api}.
For all our testing, we used a UK IP address.

In a next step, we used puppeteer, again, to open all the pages behind the first search results.
We used the popular browser extension Consent-O-Matic to get past most consent pop-ups on websites~\cite{nouwens_dark_2020}.
We also kept track of all connections to third-party URLs and checked them against the list of known third-party tracking URLs using DuckDuckGo's list of known third-party tracking domains, which is commonly used for this purpose.
We chose to study this so as to understand the privacy intrusiveness of the first search results across the different search engines, given that Google's business model vastly relies on the collection of data about individuals~--~both on and outside Google Search.

With permutation tests ($n=10,000$), we tested for statistical significant of the means in our descriptive statistics of search ranks and included trackers.

\begin{figure*}
    \centering
    
    \begin{subfigure}[b]{0.3\linewidth}
        \includegraphics[width=\linewidth]{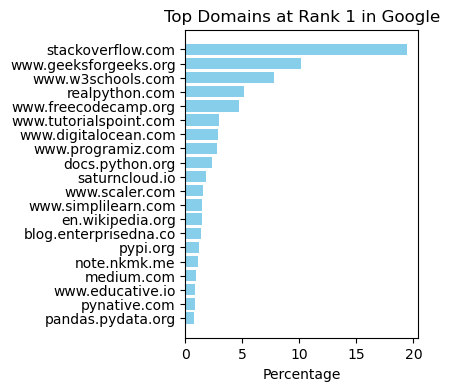}
        \caption{Top Domains at Rank 1 in Google}
        \label{fig:google}
    \end{subfigure}
    \hfill    
    \begin{subfigure}[b]{0.3\linewidth}
        \includegraphics[width=\linewidth]{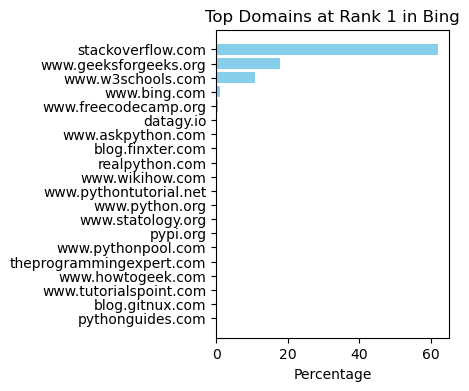}
        \caption{Top Domains at Rank 1 in Bing}
        \label{fig:bing}
    \end{subfigure}
    \hfill    
    \begin{subfigure}[b]{0.3\linewidth}
        \includegraphics[width=\linewidth]{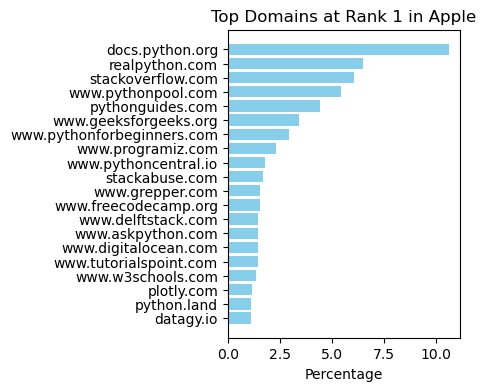}
        \caption{Top Domains at Rank 1 in Apple}
        \label{fig:apple}
    \end{subfigure}
    
    \caption{Comparative analysis of top domains at rank 1 in the given search engines.}
    \label{fig:searchengines}
\end{figure*}

\begin{figure*}
    \centering
    
    \begin{subfigure}[b]{0.3\linewidth}
        \includegraphics[width=\linewidth]{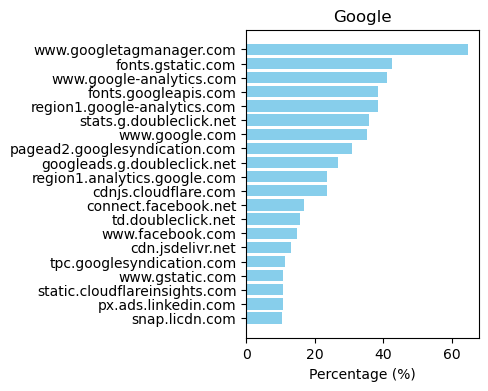}
        \caption{Top Trackers in Google}
        \label{fig:google_trackers}
    \end{subfigure}
    \hfill
    \begin{subfigure}[b]{0.3\linewidth}
        \includegraphics[width=\linewidth]{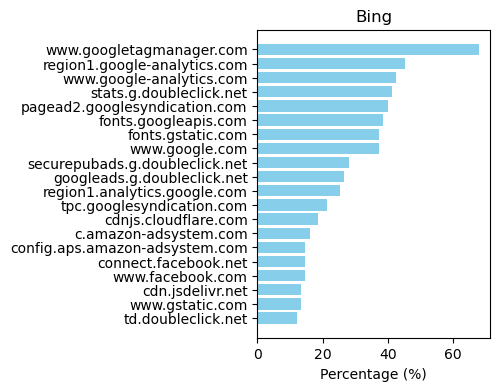}
        \caption{Top Trackers in Bing}
        \label{fig:bing_trackers}
    \end{subfigure}
    \hfill
    \begin{subfigure}[b]{0.3\linewidth}
        \includegraphics[width=\linewidth]{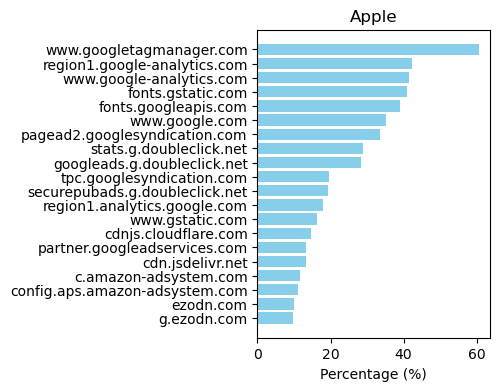}
        \caption{Top Trackers in Apple}
        \label{fig:apple_trackers}
    \end{subfigure}
    
    \caption{Most commonly contacted third-party tracking domain by the websites behind the first search results.}
    \label{fig:top_trackers}
\end{figure*}

\section{Results}
\label{sec:results}

\textbf{Most common domains at first rank.}
We first focus on the most common domains at the first rank following 1,467 queries. This is visualised in Figure~\ref{fig:searchengines}.
The search engines all reveal subtle differences. On both Google and Bing, Stack Overflow comes first, but is much more commonly at first rank Bing (61.2\%) compared to Google Search (19.4\%). On both search engines, this is followed by \url{geeksforgeeks.org} (10.2\% Google, 17.8\% Bing) and \url{w3schools.com} (7.7\% Google, 10.9\% Bing).
Both websites are well-known in the programming scene to provide general documentation on coding, albeit not very specific advice.
All other domains have less than 5\% share of the top results on Google.
Meanwhile, the most common domain on Apple Search is \url{docs.python.org}, which is the official documentation provided by the Python developers, in 10.6\% of all searches. All other domains have a smaller share than 10\%, with Stack Overflow coming third (6.1\%).

\begin{figure}
    \centering
    \begin{tabular}{lcc}
        \toprule
        \textbf{Search Engine} & \textbf{Stack Overflow, Avg. Rank} & \textbf{Avg. Tracker Number}\\ 
        \midrule
        Google Search                 & 2.8                      &  13.2                             \\ 
        Microsoft Bing                & 1.9                    &  12.1                             \\ 
        Apple Search                  & 3.6                 &  15.1                            \\ 
        \bottomrule
    \end{tabular}
    \caption{Average Rank of Stack Overflow and Average Number of Trackers on URL behind First Search Results.}
    \label{tab:ranks}
    \label{tab:trackers}
\end{figure}

\textbf{Average rank of Stack Overflow.}
We computed the average rank for Stack Overflow in the search results, see Figure~\ref{tab:ranks}. Since Apple Search only returns three results, we assigned a rank of 4 whenever Stack Overflow was not in those top three results.
Stack Overflow had an average rank of 2.8 on Google Search, 1.9 on Microsoft Bing, and 3.6 on Apple Search.
The differences between the means were pair-wise statistically significant.
The fact that the search rank on Apple Search is higher than 3 highlights that Stack Overflow does not show up in most search results in Apple Search.

\textbf{Average number of tracking domains behind first result.}
Using puppeteer and DuckDuckGo's TrackerRadar (see Section~\ref{sec:methods}), we opened the URL behind the first search results (once for each unique domain), see Figure~\ref{tab:trackers} for descriptive statistics.
Microsoft Bing results had, on average, the lowest number of tracker domains (12.1), following by Google Search (13.2) and Apple Search (15.1).
The differences between the means were pair-wise statistically significant.

\textbf{Most common tracking domains behind first result.}
Figure~\ref{fig:top_trackers} gives more details and shows bar charts of the most commonly contacted domains.
As expected, Google domain are the most commonly used third-party domain. \url{googletagmanager.com} was most commonly contacted third-party domain on about 60\% of first search results of the three studied search engines.
Google Tag Manager is a software that allows the management of `tags', which are tracking and analytics tools.
Indeed, for each search engine, the top ten most commonly contacted domains all belong to Google, including
\url{fonts.gstatic.com} (Google Fonts), \url{google-analytics.com} (Google Analytics) and \url{doubleclick.net} (Google Ads).
This underlines how Google widely collects data and shows adverts on third-party websites.

\section{Discussion}

In the study, Bing performed best for coding search quality, in terms of privacy and Stack Overflow integration.
As discussed before, Microsoft even sometimes integrates the results from Stack Overflow into the search results, thereby reducing the steps needed for developers to find suitable pieces of code. The most recent iterations of Bing even directly integrate ChatGPT, which has become the state-of-the-art in terms of retrieving quality coding advice~--~albeit not further studied in this report.
However, Microsoft's integration of ChaptGPT and also its development of the GitHub CoPilot shows its commitment to providing software developers with high-quality coding advice.
As regards the privacy findings, Stack Overflow features the integration of very few third-party services (only 6 third-party domains) and thus decreases the overall number of contacted third-party domains.

Meanwhile, Google Search shows many niche websites with coding-related advice.
If one studies the mentioned domains carefully from the perspective of a programmer (compare Figure~\ref{fig:google}), one notices many websites that are not well-known in the programming community.
Indeed, anecdotal personal evidence suggests that many of the websites that one finds through Google when looking for coding advice are cluttered with adverts and relevant information is difficult to find in long amounts of text about coding that made users spend more time on the website than necessary, as a result of SEO. This may also partly explain why so many software developers have now swiftly moved over to LLMs for coding advice because the experience on Google has long been subpar.
This subpar experience is reflected by the fact that Stack Overflow has lost up to 35\% of its usual traffic following the release of ChatGPT~\cite{stackoverflow-traffic}.

Given Google's dominance in search and advertising, there might also be reason to believe that Google, in the past, did not actually have an interest in helping users find coding-related content fast but rather have them see as many adverts as possible.
This is, for example, underlined by Sridhar Ramaswamy quitting as the head of Google's advertising division over concerns that Google was not giving sufficient priority to the user experience in search but only focused on showing them more ads and collecting more data about them~\cite{neeva}.
In response, he chose to co-found Neeva in 2019, an alternative search engine that came with a subscription model instead of ads. Given the high competitive pressure and barriers to entry into search, the company ultimately was disbanded in 2023.

Lastly, Apple Search is relatively nascent and rarely used for coding-related queries.
In our results, Apple Search performed worst in terms of search quality.
This would be expected, given that it is the least widely used search engine and sees much fewer queries than Google or Bing.

\section{Conclusions and suggestions for future work on search quality}

We presented a quantitative comparison of search quality, using the case study of coding advice.
To this end, we compared the search quality on Google Search, Microsoft Bing and Apple Search. Importantly, Apple Search had rarely been studied or documented before, and this report is the first to look into this part of the ecosystems.
Under the assumption that Stack Overflow usually gives the best coding advice, our results suggest that the search quality, when it comes to coding advice, is significantly higher on Bing as compared to Google and Apple, and that the use of Bing has less implications for privacy.

Given that search quality has rarely been assessed over time or at scale, we think this needs to change and make suggestions as to how this could be achieved.

\textbf{Develop and define robust criteria for search quality.} For sector-specific search, this can be the rank of Stack Overflow, as done in this work. For general search, this can include the third-party tracking and privacy invasiveness of the first search results, as also done in this work. Moving forward, one may want to measure the time it takes an individual until they have gathered and found the information that they need. If a user is logged into an account with a search providers, the semantic assessment of the length of users' sessions and time to find information can likely be approximated. Such an analysis would not only be possible on traditional search but also for LLM platforms like ChatGPT.

\textbf{Make full use of legal mechanisms to retrieve search data.} In order to get detailed insights into search quality, the EU's Digital Markets Act (DMA) is particularly promising. Under this law, digital gatekeepers have to fulfil a stringent list of obligations as regards their core services.
Google Search, unlike Bing, is one of those core services of a digital gatekeeper (Google) and must follow those obligations.
Under Article 6(11) DMA, Google has to provide `to any third-party undertaking providing online search engines, at its request, with access on fair, reasonable and non-discriminatory terms to ranking, query, click and view data' regarding Google Search.
In the future, there is a chance that platforms like ChatGPT might be covered, too, as soon as the platform rises above the threshold of 45 million monthly active users, as is required under the DMA.
Additionally, the EU's Digital \textit{Services} Act (DSA) has further provisions on researcher data access under Article 40 DSA in relation to very large online platforms and search engines, including Google Search.

\textbf{Continuously collect data and make it public.}
The last major large-scale study into search quality was conducted based on 2011 data, which was before Google made many major changes to its search infrastructure~\cite{lewandowski_evaluating_2015}.
Thus, we remain in the dark about how search quality has developed over time and how it is doing today.
As more individuals shift to LLMs which are arguably even more opaque than traditional search, such an assessment of search/information quality will be ever more important.
Making insights into search quality public will create natural pressure on companies to be held accountable and to improve search quality over time, which is fundamental for our information society.

\begin{figure}
\centering
\begin{lstlisting}
const response = await axios.get(
  'https://api-glb-aeuc1a.smoot.apple.com/search', {
  params: {
    '24h': '1',
    'alwaysSendTophit': 'on',
    'calendar': 'Gregorian Calendar',
    'cc': 'GB',
    'esl': 'en',
    'kb_ime': 'gb',
    'key': 'puffin1758',
    'l2': '13.0.72',
    'locale': 'en_DE',
    'magic': 'bullseye,bullseye2',
    'q': 'python how to download a file from url',
    'qtype': 'suggest_with_results',
    'siri_locale': 'en-GB',
    'storefront': '143444,29',
    'temp': 'C',
    'time_zone': 'Europe/London',
    'units': 'SI'
  },
  headers: {
    'Host': 'api-glb-aeuc1a.smoot.apple.com',
    'x-apple-ui-scale': '1.0',
    'accept': '*/*',
    'accept-language': 'en-GB,en;q=0.9',
    'x-apple-subscriptions': '[]',
    'x-apple-locage': '<15m',
    'user-agent': 'parsecd/1 (MacBookAir10,1; macOS 14.0 23A344) spotlight/1',
    'x-apple-languages': '["en-GB"]',
    'x-apple-seed': '75',
    'x-apple-geocountrycodesource': 'geoIP'
  }
});
\end{lstlisting}
\caption{An example request to Apple Search with the query `python how to download a file from url'.
This API has hardly previously been documented.
}
\label{fig:apple-api}
\end{figure}

\bibliographystyle{ieeetr}
\bibliography{references}

\end{document}